\documentclass[notitlepage,reqno,a4paper]{article}
\usepackage{}
\usepackage{amsfonts}
\usepackage{subfigure}
\usepackage{cite}
\usepackage{mathrsfs}
\usepackage{stmaryrd}
\usepackage{amsmath}
\usepackage{amssymb}
\usepackage{graphicx}
\usepackage{verbatim}
\usepackage{indentfirst}
\usepackage{txfonts}

\begin{document}
\begin{center}\large \textbf{L\'evy path integral approach to the solution of the fractional Schr\"odinger equation with infinite square well}
\end{center}
\begin{center}\small {Jianping Dong\footnote{Email:Dongjp.sdu@gmail.com}}\\
{\itshape Department of Mathematics, College of Science, Nanjing
University of Aeronautics and Astronautics, Nanjing 210016,
China}\end{center}
\begin{quote}  The solution to the fractional Schr\"odinger equation with infinite square well is obtained in this paper, by use of the L\'evy path integral approach. We obtain the even and odd parity wave functions of this problem, which are in accordance with those given by Laskin in [Chaos 10 (2000), 780--790].\end{quote}
\section{{Introduction}}
\label{sec1}
The phrase  \lq\lq fractional quantum mechanics \rq\rq was first introduced by Nick Laskin \cite{laskin1,laskin2}. The fractional quantum system is described
 by the fractional Schr\"odinger equation (FSE) \cite{laskin3} containing the fractional Riesz operator \cite{kilbas}. It is well-known that Feynman and Hibbs \cite{feynman} reformulated the standard Schr\"odinger equation \cite{levin} by use of the path integral  approach considering the Gaussian probability distribution. The L\'evy stochastic process is a natural generalization of the Gaussian process.
The possibility of developing the path integral over the paths of the L\'evy motion was
discussed by Kac \cite{kac}, who pointed out that the L\'evy path
integral generates the functional measure in the space of left (or
right) continued functions having only discontinuities of the first
kind. Recently, Laskin\cite{laskin1} generalized
Feynman path integral to L\'evy one, and developed the
space-fractional Schr\"odinger equation containing the Riesz
fractional derivative \cite{laskin3,kilbas}. Then, he constructed
the fractional quantum mechanics and showed some properties of the
space fractional quantum system \cite{laskin2,laskin4,laskin5}. Afterwards, the time-fractional Schr\"odinger equation containing the Caputo fractional derivative \cite{kilbas}
, and  the space-time-fractional Schr\"odinger
equation \cite{naber,wang,dong1} were also given. Because of the double complexities of the fractional calculus and quantum mechanics, not too much progress has been made in this field. The solutions of the FSE with the linear potential, the delta potential,   the infinite square well, and so on, have been obtained \cite{laskin4,laskin5 ,dong2,dong3,guo,Oliveira}. The fractional scattering problem for the one- and three-dimensional cases were also studied \cite{dong4,dong5}, and the Thomas-Fermi theory for the multi-particle system was also generalized to the fractional quantum system \cite{dong6}.
 \par This paper focuses on the solution to the fractional  Schr\"odinger equation with infinite square well. This solution was firstly obtained by Laskin in the piece-wise
solution approach, but then in Ref.~\cite{Jeng}, Jeng, eta. pointed out that this solution and some other obtained solutions to the fractional Schr\"odinger equation  with specific potentials were incorrect, because of the nonlocal property of the fractional Riesz operator. Then, in Ref.~\cite{Bayin}, Bayin argued that the proof given by Jeng, eta., that the solution gotten by Laskin did not satisfy the FSE, was not true, and  proved Laskin's solution to be correct. But the controversy on this point still do not stop yet, see, for example, Ref.~\cite{Hawkins}. In this paper, we attempt to get the solution of the FSE with infinite square well in an indirect way: the L\'evy  path integral approach.

\section{{L\'evy  path integral and  Fractional Schr\"odinger Equation}} \label{sec1}
\label{sec2}The L\'evy  path integral is generalized from the Feynman one by Laskin in Refs.~\cite{laskin1,laskin5}. In this section, we give the introduction to the L\'evy  path integral and  fractional Schr\"odinger equation, and show some properties useful in the following sections. If a particle at an initial time $t_a$ starts from the point $x_a$ and goes to a final point $x_b$ at time $t_b$, we will say simply that the particle goes from to $a$ and $b$ and its trajectory (path) $x(t)$ will have the property that $x(t_a)=x_a$ and $x(t_b)=x_b$. In quantum mechanics, then, we will have an quantum-mechanical amplitude, often called a kernel, which we may write $K(x_bt_b|x_at_a)$, to get from the point $a$ to the point $b$. This will be the sum over all of the paths that go between that end points and of a contribution from each. If we have the quantum particle moving in the potential then the quantum-mechanical amplitude $K_L(x_bt_b|x_at_a)$ can be written as
\begin{equation}
\begin{split}
K_L(x_bt_b|x_at_a)=\int_{x(t_a)=x_a}^{x(t_b)=x_b}\mathscr{D}x(\tau)\cdot \exp\left\{-\frac{i}{\hslash}\int_{t_a}^{t_b}d\tau V(x(\tau))\right\},\label{kernel1}
\end{split}
\end{equation}where $V(x(\tau))$ is the potential energy as a functional of the L\'evy particle path and the fractional path integral measure is defined as
\begin{equation}
\begin{split}
\int_{x(t_a)=x_a}^{x(t_b)=x_b}\mathscr{D}x(\tau)=\lim_{N\rightarrow\infty}\int dx_1\cdots dx_{N-1}\hslash^{-N}\left(\frac{iD_\alpha\varepsilon}{\hslash}\right)^{-N/\alpha}\cdot\prod_{j=1}^NL_\alpha\left\{\frac1\hslash
\left(\frac{\hslash}{iD_\alpha\varepsilon}\right)^{1/\alpha}|x_j-x_{j-1}|\right\}\cdots,
\end{split}
\end{equation} where $D_\alpha$ is the generalized 'fractional diffusion coefficient', the physical
dimension of which is $[D_\alpha]=\text{erg}^{1-\alpha}\times \text{cm}^\alpha
\times \text{sec}^{-\alpha}$ ($D_\alpha=1/2m$ for $\alpha=2$, $m$ denotes the mass of a particle)
, $x_0=x_a$, $x_N=x_b$, $\varepsilon=(t_b-t_a)/N$, and the L\'evy probability distribution function $L_\alpha$ is expressed in term of Fox's H function (see \cite{laskin1}).
The free particle kernel of the L\'evy type can be expressed in the momentum representation by
\begin{equation}
\begin{split}
K_L^{(0)}(x_bt_b|x_at_a)=\frac{1}{2\pi\hslash}\int_{-\infty}^{+\infty}dp\cdot\exp\left\{i\frac{p(x_b-x_a)}{\hslash}-i\frac{D_\alpha(t_b-t_a)|p|^\alpha}{\hslash}\right\}.
\label{freepathintegral}
\end{split}
\end{equation}
The kernel $K_L(x_bt_b|x_at_a)$ which is defined by Eq.~(\ref{kernel1}), describes the evolution of the fractional quantum-mechanical system,
\begin{equation}
\begin{split}
\psi_f(x_b,t_b)=\int_{-\infty}^{+\infty}dx_aK_L(x_bt_b|x_at_a)\psi_i(x_b,t_b),\label{waverelation1}
\end{split}
\end{equation}
where $\psi_i(x_b,t_b)$ is the fractional wave function of initial (at the $t=t_a$)state and
 $\psi_f(x_b,t_b)$ is the fractional wave function of the final (at the $t=t_b$) state. With the help of this equation, for the fractional quantum system, Laskin derived the fractional Schr\"odinger equation given as follows (in one-dimensional case),
\begin{equation}
i\hslash\frac{\partial\psi(x,t)}{\partial t}=H_\alpha \psi(x,t),
\label{fseeq1}
\end{equation}
where $\psi(x,t)$ is the time-dependent wave function , and
$H_\alpha(1<\alpha\leq 2)$ is the fractional Hamiltonian operator
given by
\begin{equation}
H_\alpha=-D_\alpha (\hslash\nabla)^{\alpha}+V(x,t).
\label{fseeq2}
\end{equation}
Here, $(\hslash\nabla)^{\alpha}$ is the quantum Riesz
fractional operator defined by
\begin{equation}
(\hslash\nabla)^{\alpha}\psi(x,t)=-\frac1{2\pi\hslash}\int^{+\infty}
_{-\infty}\mathrm{d}p\text{e}^{ipx/\hslash}|p|^\alpha\int^{+\infty}_{-\infty}\text{e}^
{-ipx/\hslash}\psi(x,t)\mathrm{d}x \label{fseeq3}.
\end{equation}
In Ref.,  the  fractional Hamiltonian operator was proved to be a Hermitian or self-adjoint operator. When $H_\alpha$ is not explicitly dependent on $t$, that is, $V(x,t)=V(x)$, after separation of variables, Eq. (\ref{fseeq1}) goes over to a form of steady state,
\begin{equation}
-D_\alpha (\hslash\nabla)^{\alpha}\phi(x)+V(x)\phi(x)=E\phi(x),
\label{fseeq4}
\end{equation}
where $\psi(x,t)$ and $\phi(x)$ are related to each other by $\psi(x,t)=\phi(x)\text{e}^{-iEt/\hslash}$ ,
in which $E$ denotes the energy of the quantum system. Suppose that the solutions $\phi_n(x)$ (time-independent wave functions) to Eq.~(\ref{fseeq4}) corresponding to the set of energy levels $E_n$ are not only orthogonal but also normalized, then following Fenyman's method, we can obtain the relationship between the fractional quantum-mechanical kernel and the time-independent wave functions $\phi_n(x)$,
\begin{equation}
K_L(x_bt_b|x_at_a)=\sum_{n=1}^{\infty}\phi_n(x_b)\phi_n^*(x_a)e^{-(i/\hslash)E_n(t_b-t_a)},\mbox{{ }  for { } } t_b>t_a \label{kwrelation}
\end{equation} and $K_L(x_bt_b|x_at_a)=0$, for $t_b<t_a$; $K_L(x_bt_b|x_at_a)=\delta(x_b-x_a)$, for $t_b=t_a$. Eq.~(\ref{kwrelation}) will be used in the following section to solve the fractional Schr\"odinger equation with infinite square well.

\section{{Infinite square well: wave functions and energy eigenvalues}}
\label{sec3}In this section, we consider a particle in the infinite square well with the potential $V(x)$ defined as
\begin{equation} V(x)=\left\{
\begin{aligned}0, &&&|x|\geq l, \\\infty, &&&|x|> l.
\end{aligned} \right.
\label{potentialf}
\end{equation} The solution to the fractional Schr\"odinger equation with this potential was firstly studied by Laskin in Ref.~\cite{laskin4}, in which this equation is solved in a piecewise approach. In this paper, to avoid  treating the nonlocal property of the fractional Riesz operator, we get the solutions to this equation in the L\'evy path
integral approach using Eq.~(\ref{kwrelation}). That is, we turn to get the fractional quantum-mechanical kernel of this quantum system. Now we use the path cancellation method to get the transfer kernel.
 \par An interpretation of the infinite potential barrier problem was the observation that there are two classical paths connecting the points $(x_a,t_a)$ and $(x_b,t_b)$ \cite{Goodman}. This is because the classical particle may go directly between the two points, or it may bounce off the wall once on the way. With the infinite well, the particle may bounce back and forth an arbitrary number of times on its way from $(x_a,t_a)$ to $(x_b,t_b)$. This leads to an infinite number of classical paths connecting the two points. These paths correspond to an infinite sequence of images of the final point having coordinates
\begin{equation} x_r=\left\{
\begin{aligned} 2lr+x_b, &&&\mbox{ when }r=2m, \\2lr-x_b, &&&\mbox{ when }r=2m+1,
\end{aligned} \right.
\label{coordinates}
\end{equation}in which $m=0,\pm1,\pm2,\cdots$. Here, the subscript $r$ is used to denote the number of reflections in the classical path corresponding to the image point, and to distinguish from the subscript $n$ used for the energy eigenstates.
\par A generalization of the above arguments for the potential barrier would suggest that the propagator $K_L(x_bt_b|x_at_a)$ should be the sum of contributions from each of these classical paths. The contribution from each classical path should be the free-particle propagator for its corresponding image point, multiplied by $-1$ for each time it is reflected. This leads to the kernel,
\begin{equation}
K_L(x_bt_b|x_at_a)=\sum_{r=-\infty}^{+\infty}(-1)^rK_L^{(0)}(x_rt_b|x_at_a)=\sum_{m=-\infty}^{+\infty}[K_L^{(0)}(x_{2m}t_b|x_at_a)-K_L^{(0)}(x_{2m+1}t_b|x_at_a)].
\label{kernel1}
\end{equation}
Recalling Eqs.~(\ref{freepathintegral}) and~(\ref{coordinates}), the kernel can be calculated as follows,
\begin{equation}
\begin{split}
&K_L(x_bt_b|x_at_a)=\sum_{m=-\infty}^{+\infty}\frac{1}{2\pi\hslash}\int_{-\infty}^{+\infty}
\left\{\exp\left[i\frac{p(4ml+x_b-x_a)}{\hslash}-i\frac{D_\alpha(t_b-t_a)|p|^\alpha}{\hslash}\right]-
\exp\left[i\frac{p(4ml+2l-x_b-x_a)}{\hslash}-i\frac{D_\alpha(t_b-t_a)|p|^\alpha}{\hslash}\right]\right\}dp
\\&=\frac{1}{2\pi\hslash}\int_{-\infty}^{+\infty}\exp\left[-i\frac{D_\alpha(t_b-t_a)|p|^\alpha}{\hslash}\right]
\left\{\exp\left[i\frac{p[(x_b-l)-(x_a-l)]}{\hslash}\right]-
\exp\left[i\frac{p[-(x_b-l)-(x_a-l)]}{\hslash}\right]\right\}\sum_{m=-\infty}^{+\infty}\exp\left[i\frac{4mlp}{\hslash}\right]dp
\\&=\frac{i}{\pi\hslash}\int_{-\infty}^{+\infty}\exp\left[-i\frac{D_\alpha(t_b-t_a)|p|^\alpha}{\hslash}\right]\exp\left[i\frac{-p(x_a-l)}{\hslash}\right]
\sin(\frac{p(x_a-l)}{\hslash})\sum_{m=-\infty}^{+\infty}\exp\left[i\frac{4mlp}{\hslash}\right]dp.
\end{split}
\label{kernelcal1}
\end{equation}
Furthermore, using the Poisson summation rule
\begin{equation}
\sum_{r=-\infty}^{+\infty}\exp(2\pi im\mu)=\sum_{n=-\infty}^{+\infty}\delta(\mu-n),
\label{poisson}
\end{equation}
Then, we have
\begin{equation}
\begin{split}
&K_L(x_bt_b|x_at_a)=\frac{i}{\pi\hslash}\int_{-\infty}^{+\infty}\exp\left[-i\frac{D_\alpha(t_b-t_a)|p|^\alpha}{\hslash}\right]\exp\left[i\frac{-p(x_a-l)}{\hslash}\right]
\sin(\frac{p(x_a-l)}{\hslash})\sum_{n=-\infty}^{+\infty}\delta(\frac{2lp}{\pi\hslash}-n)dp
\\&=\frac{i}{2l}\sum_{n=-\infty}^{+\infty}\int_{-\infty}^{+\infty}\exp\left[-i\frac{D_\alpha(t_b-t_a)|p|^\alpha}{\hslash}\right]\exp\left[i\frac{-p(x_a-l)}{\hslash}\right]
\sin\left(\frac{p(x_a-l)}{\hslash}\right)\delta\left(p-\frac{n\pi\hslash}{2l}\right)dp
\\&=\frac{i}{2l}\sum_{n=-\infty}^{+\infty}\exp\left[-iE_n(t_b-t_a)\right]\exp\left[-ik_n(x_a-l)\right]
\sin\left[k_n(x_a-l)\right],
\end{split}
\label{kernelcal2}
\end{equation} in which $$k_n=\frac{n\pi}{2l},E_n=D_\alpha\hslash^\alpha|k_n|^\alpha.$$
Noting that the term with $n=0$ vanishes, we can combine the positive and negative terms to get
\begin{equation}
\begin{split}
&K_L(x_bt_b|x_at_a)=\frac{1}{l}\sum_{n=1}^{\infty}\exp\left[-iE_n(t_b-t_a)\right]\sin\left[k_n(x_a-l)\right]
\sin\left[k_n(x_a-l)\right].
\end{split}
\label{kernelcal3}
\end{equation}
Comparing Eq.~(\ref{kernelcal3}) with Eq.~(\ref{kwrelation}) yields the wave functions and energy levels,
\begin{equation}
\phi_n(x)=\frac{1}{\sqrt{l}}\sin\left[k_n(x-l)\right]=\frac{1}{\sqrt{l}}\sin\left[\frac{n\pi}{2l}(x-l)\right],
 E_n=D_\alpha\left(\frac{n\pi\hslash}{2l}\right)^\alpha,n=1,2,3,\cdots.
\label{solution1}
\end{equation}
These results seem different from those given by Laskin. In fact, we have
\begin{equation}
\phi_n(x)=\left\{
\begin{aligned}\frac{1}{\sqrt{l}}\sin\left(\frac{j\pi}{l}x-j\pi\right)=\frac{(-1)^j}{\sqrt{l}}\sin\left(\frac{j\pi}{l}x\right),\qquad &&&\mbox{ when } n=2j, j=1,2,3,\cdots
\\\frac{1}{\sqrt{l}}\sin\left(\frac{j\pi}{l}x-j\pi-\frac{\pi}{2}\right)=\frac{(-1)^{j+1}}{\sqrt{l}}\cos\left(\frac{j\pi}{l}x\right). &&&\mbox{ when } n=2j+1,j=0,1,2,3,\cdots
\end{aligned} \right.
\label{solution2}
\end{equation}
Omitting the product factor, we can get the even and odd parity wave functions and the corresponding energy levels,
\begin{equation}
\begin{split}
&\phi_n^{\mbox{odd}}=\frac{1}{\sqrt{l}}\sin\left(\frac{n\pi}{l}x\right),E_n=D_\alpha\left(n\frac{\pi\hslash}{l}\right)^\alpha;
\\&\phi_n^{\mbox{even}}=\frac{1}{\sqrt{l}}\cos\left(\frac{n\pi}{l}x\right),E_n=D_\alpha\left((n+\frac12)\frac{\pi\hslash}{l}\right)^\alpha,
\end{split}
\end{equation}which are in accordance with the results given by Laskin.

\section{{Conclusions}}
In this paper, we solved the fractional  Schr\"odinger equation with infinite square well in the L\'evy path integral approach. Because of the nonlocal property of the fractional Riesz operator, the usual method for solving the standard Schr\"odinger equation can not be effective here. The solution to the equation gotten by the usual method has been proved to be true, but weather the usual method is right or not, or when the usual method is effective, still need more investigations. The method given in this paper, can be a way to solve the  fractional  Schr\"odinger equation indirectly.

\section*{{Acknowledgements}}
This work was supported by the National Natural Science Foundation of China (Grant
No. 11147109), the Specialized Research Fund for the Doctoral Program of Higher Education
of China (Grant No. 20113218120030), and the Fundamental Research Funds for the Central
Universities (Grant No. NS2012119).


\footnotesize
\begin{thebibliography}{33}
\bibitem{laskin1} N. Laskin, Fractional quantum mechanics and L\'evy path integrals,
Phys. lett. A 268 (2000) 298--305.
\bibitem{laskin2} N.Laskin, Fractional quantum mechanics, Phys. Rev. E 62 (2000) 3135--3145.
\bibitem{laskin3} N.Laskin, Fractional Schrodinger equation, Phys. Rev. E 66 (2002) 056108.
\bibitem{kilbas} A.A. Kilbas, H.M. Srivastava and J. J. Trujillo, Theory and Applications of
Fractional Differential Equations, Elsevier, Amsterdam, 2006.

\bibitem{feynman} R.P. Feynman and A. R. Hibbs,
Quantum Mechanics and path Integrals, McGraw-Hill, New York,1965.

\bibitem{levin} F.S. Levin, An Introduction to Quantum Theory, Cambridge University
Press, 2002.

\bibitem{kac} M. Kac, On Some Connections between Probability Theory and Differential and Integral
Equations, in Second Berkeley Symposium on Mathematical Statistics
and Probability (edited by Jerzy Neyman), University of California
Press, Berkeley, California, 1951.

\bibitem{laskin4} N.Laskin, Fractals and quantum mechanics, Chaos 10 (2000) 780--790.

\bibitem{laskin5} N.Laskin, Levy flights over quantum paths, Comm. Nonlin. Sci. Num. Sim. 12(2), 2007.

\bibitem{naber} M. Naber, Time fractional
Schrodinger equation, J. Math. Phys. 45(8), 2004. 3339-3352.
\bibitem{wang} S.W. Wang and M.Y. Xu, Generalized fractional Schrodinger
equation with space-time fractional derivatives, J. Math. Phys. 48
(2007) 043502.
\bibitem{dong1} J.P. Dong, M.Y. Xu, Space-time fractional
Schrodinger equation with time-independent potentials, J. Math.
Anal. Appl., 344, 1005-1017 (2008).

\bibitem{dong2} J.P. Dong, M.Y. Xu, Applications of continuity
and discontinuity of a fractional  derivative of the wave functions
to fractional quantum mechanics, J. Math. Phys. 49, 052105 (2008).
\bibitem{dong3} J.P. Dong, M.Y. Xu, Some solutions to
the space fractional Schrodinger equation using momentum
representation method, J. Math. Phys. 48 (2007) 072105.
\bibitem{guo} X.Y. Guo, M.Y. Xu, Some physical applications of fractional Schrodinger equation, J. Math. Phys. 47(2006) 082104.

\bibitem{Oliveira} E. C. Oliveira,  F.S. Costa and J. J. Vaz, The fractional Schr?dinger equation for delta potentials, J. Math. Phys. 51, 123517 (2010).

\bibitem{dong4} J. P. Dong,  Generalized Lippmann-Schwinger equation in the fractional quantum mechanics, Journal of Physics A: Mathematical and Theoretical, 44 (2011) 215204;

\bibitem{dong5} J. P. Dong, Green¡¯s function for the time-dependent scattering problem in the fractional quantum mechanics, Journal of Mathematical physics, 52 (2011) 042103 ;
    
\bibitem{dong6} J. P. Dong, Applications of density matrix in the fractional quantum mechanics: Thomas-Fermi model and Hohenberg-Kohn theorems revisited, Physics Letters A, 375 (2011) 2787-2792.

\bibitem{Jeng} M. Jeng, S.-L.-Y. Xu, E. Hawkins, and J. M. Schwarz, J. Math. Phys. 51, 062102 (2010).

\bibitem{Bayin} S. S. Bayin, On the consistency of the solutions of the space fractional
Schr\"odinger equation, J. Math. Phys. 53, 042105 (2012).

\bibitem{Hawkins} E. Hawkins and J. M. Schwarz, Comment on: \lq\lq the consistency of solutions of the space fractional Schr\"odinger equation\rq\rq, arXiv: 1210.1447.

\bibitem{Goodman} M. Goodman, Path integral solution to the infinite square well, American Journal of Physics, 49(1981), 843.


\end{thebibliography}
\end{document}